\documentclass[aps, twocolumn, letterpaper, superscriptaddress, nofootinbib]{revtex4}

\usepackage{amsmath}
\usepackage{amssymb}
\usepackage{mathrsfs}
\usepackage{xspace}
\usepackage{braket}
\usepackage{color}
\usepackage{graphicx}

\usepackage{ifpdf}
\ifpdf
\fi

\newcommand{\myRef}[1]{Ref.~\cite{#1}}
\ifx\Ref\undefined
\newcommand{\Ref}[1]{\myRef{#1}}
\else
\renewcommand{\Ref}[1]{\myRef{#1}}
\fi

\newcommand{\eq}[1]{(\ref{#1})}
\newcommand{\Eq}[1]{Eq.~(\ref{#1})}
\newcommand{\Eqs}[1]{Eqs.~(\ref{#1})}

\newcommand{\App}[1]{Appendix~\ref{#1}}
\newcommand{\Sec}[1]{Sec.~\ref{#1}}
\newcommand{\Refs}[1]{Refs.~\cite{#1}}

\newcommand{\eg}{{e.g.,\/}\xspace}
\newcommand{\ie}{{i.e.,\/}\xspace}
\newcommand{\etal}{{\it et~al.\/}\xspace}

\newcommand{\pd}{\partial}
\newcommand{\dd}{\mathrm{d}}
\newcommand{\ee}{\mathrm{e}}
\newcommand{\ii}{\mathrm{i}}
\newcommand{\del}{\nabla}

\newcommand{\mc}[1]{\mathcal{#1}}
\newcommand{\mcc}[1]{\mathfrak{#1}}
\newcommand{\msf}[1]{\mathsf{#1}}

\renewcommand{\vec}[1]{{\boldsymbol{#1}}}
\newcommand{\oper}[1]{\smash{\hat{#1}}}
\newcommand{\boper}[1]{\oper{\vec{#1}}}
\newcommand{\kpt}[1]{\kern #1 pt} 
\newcommand{\favr}[1]{\langle #1 \rangle} 

\newcommand{\unitoper}{\oper{I}}
\newcommand{\heff}{\hbar'}

\begin{document}

\title{On applications of quantum computing to plasma simulations}

\author{I. Y. Dodin}
\affiliation{Princeton Plasma Physics Laboratory, Princeton, NJ 08543}
\affiliation{Department of Astrophysical Sciences, Princeton University, Princeton, NJ 08544}
\author{E. A. Startsev}
\affiliation{Princeton Plasma Physics Laboratory, Princeton, NJ 08543} 

\date{\today}

\begin{abstract}
Quantum computing is gaining increased attention as a potential way to speed up simulations of physical systems, and it is also of interest to apply it to simulations of classical plasmas. However, quantum information science is traditionally aimed at modeling linear Hamiltonian systems of a particular form that is found in quantum mechanics, so extending the existing results to plasma applications remains a challenge. Here, we report a preliminary exploration of the long-term opportunities and likely obstacles in this area. First, we show that many plasma-wave problems are naturally representable in a quantumlike form and thus are naturally fit for quantum computers. Second, we consider more general plasma problems that include non-Hermitian dynamics (instabilities,  irreversible dissipation) and nonlinearities. We show that by extending the configuration space, such systems can also be represented in a quantumlike form and thus can be simulated with quantum computers too, albeit that requires more computational resources compared to the first case. Third, we outline potential applications of hybrid quantum--classical computers, which include analysis of global eigenmodes and also an alternative approach to nonlinear simulations.
\end{abstract}

\maketitle

\sloppy

%\bibliographystyle{full}
%%%%%%%%%%%%%%%%%%%%%%%%%%%%%%%%%%%%%%%
\section{Introduction}
\label{sec:intro}

Recently, quantum computing (QC) has been gaining increased attention as a potential way to significantly speed up simulations of physical systems \cite{ref:montanaro16}. The focus is usually made on modeling many-body quantum systems, whose enormous configuration space is often straightforward to map on reasonably sized quantum circuits, at least in principle. But it is also of interest to explore whether QC can be useful for modeling classical systems such as plasmas. In particular, this could benefit fusion science, which heavily relies on simulations.

To assess the potential utility of QC for plasma physics, it is important to understand what a quantum computer can and cannot do naturally. A digital quantum computer usually stores information in some $N$ entangled qubits, which are two-level quantum systems. (Sometimes, $d$-level quantum systems, or qudits, are used instead, with $d > 2$.) Due to their entanglement, the total configuration space of the computer is a tensor product of the configuration spaces of individual qubits; \ie the computer state is described by a $2^N$-dimensional complex vector, $\Psi$. The exponential scaling of $\dim \Psi$ with $N$ can be advantageous in solving large-dimensional problems; however, a quantum computer is naturally fit to perform simulations only of a certain type. A quantum simulation consists of applying a sequence of a $M$ linear unitary operations (``gates'') to qubits, which results in linear unitary evolution of $\Psi$. Hence, a program is a circuit, and the practicality of a quantum simulation depends on how large $M$ and $N$ are, as these numbers are constrained by technology and, ultimately, by the computer price. Simulation results are output through classical measurements. With enough measurements, one can calculate the expectation value of any given operator on~$\Psi$ with a pre-defined accuracy, assuming that this operator is efficiently computable.

Such architecture is particularly suitable, among other things (discussed in \Sec{sec:subdiscussion} and further), to simulating processes governed by a linear Schr\"odinger equation
\begin{gather}\label{eq:schr}
\ii \pd_t \psi = \oper{H} \psi, \quad \oper{H}^\dag = \oper{H},
\end{gather}
where $\psi$ is a (quantum or classical) state vector that characterizes the physical system and the Hermitian operator $\oper{H}$ serves as a Hamiltonian. This is understood as follows. The solution of \Eq{eq:schr} is $\psi(t) = \oper{U}\psi_0$, where $\psi_0$ is the initial value of~$\psi$, $\oper{U} = \msf{T} \exp[-\ii \smash{\int_0^t} \oper{H}(t')\,\dd t']$ is a unitary evolution operator, and $\msf{T}\exp(\ldots)$ is an ordered exponential; or simply $\oper{U} = \ee^{-\ii \oper{H} t}$, if $\oper{H}$ is independent of time. A quantum circuit that implements~$\oper{U}$ can perform a \textit{quantum Hamiltonian simulation} (QHS) to yield $\psi(t)$ for given $\psi_0$. It has been shown that QHSs can be much faster than classical simulations if $\oper{H}$ is efficiently computable and of a certain type, for example, if it is represented by a sparse matrix. (For example, see the pioneering \Ref{ref:lloyd96}, the recent works \cite{ref:childs18, foot:gilyen}, and the many papers cited therein.) This also makes QHSs potentially attractive as elements of more elaborate algorithms for solving general linear equations \cite{ref:harrow09, ref:berry14, ref:childs17}. 

So far, research in this area has been focused mainly on expanding the class of Hamiltonians for which efficient QHSs are possible in an \textit{ad~hoc} fashion and on solving, basically, random problems, albeit impressively \cite{ref:arute19}. QHS implementations for problems of practical interest are rarely considered, and applicability of the existing methods to practical simulations remains uncertain \cite{ref:scherer17, ref:montanaro16b}. This is even more the case with applications of QC to non-Hermitian Hamiltonians and to nonlinear problems, which are approached \cite{ref:motta19, ref:candia15, tex:leyton08} in ways that are unlikely to benefit simulations of classical systems like plasmas. In this situation, the conceptual aspects of plasma simulations on a quantum computer need to be developed from scratch.

Here, we report a preliminary exploration of the (most obvious) long-term opportunities and likely obstacles for quantum simulations of classical plasmas. Our take on this problem is different from that of the authors who focus on quantum circuits for toy models \cite{ref:engel19, tex:shi20}. Toy models are of interest in quantum many-body physics, where even simple (efficiently mappable to qubits) Hamiltonians can produce dynamics that is both interesting and hard to simulate \cite{ref:georgescu14}. In plasma physics, though, the needs of toy-model simulations are typically satisfied already with classical computing (homogeneous turbulence may be an exception), so QC is of interest primarily for concrete practical applications. Hence, elaborate tricks that work only for special cases may not benefit the field in the long run. QC can become advantageous in plasma applications only if it can handle realistic, non-sparse, and, better yet, nonlinear Hamiltonians. Thus, rather than showing that QC can excel \textit{ad~hoc}, it may be more important to identify \textit{regular} high-level methods for mapping \textit{typical} plasma simulations on a quantum computer.

This is the problem that we address below. Because practical plasma simulations are impossible with the minimalistic quantum computers that exist today, the approaches to be discussed are intended for future universal computers with error correction \cite{ref:devitt13}. Hopefully, the elementary algorithms of QC that make it promising (\eg sparse-matrix inversion \cite{ref:harrow09, ref:clader13}) will be made reliable enough by the time when error-corrected machines appear; hence, we are not concerned with low-level building blocks of QC. This approach is justified because our paper is not about quantum algorithms \textit{per~se}; rather, it is about reducing plasma problems to QC problems.

Our paper is organized as follows. In \Sec{sec:linear}, we discuss the possibility of linear plasma simulations, particularly, related to modeling of radiofrequency (RF) waves. Both conservative and dissipative waves are considered. In \Sec{sec:nonlin}, we outline principles of nonlinear simulations in application to general dynamical systems, Hamiltonian dynamics, and fluid simulations in particular. In \Sec{sec:eigen}, we discuss QC applications to finding linear plasma eigenmodes, for example, magnetohydrodynamic modes (MHD), using hybrid quantum--classical computing. In \Sec{sec:varia}, we discuss applications of hybrid computing to nonlinear simulations. In \Sec{sec:conc}, we summarize our main results. In \App{app:llg}, we present a supplementary discussion to accentuate some of the basic ideas introduced in the main text. In \App{app:delta}, we elaborate on the definitions of the generalized functions that are used in the main text.

%%%%%%%%%%%%%%%%%%%%%%%%%%%%%%%%%%%%%%%
\section{Linear dynamics}
\label{sec:linear}

First, let us discuss the possibility of linear plasma simulations, particularly RF-wave modeling. RF waves are commonly used as precision tools, for example, for plasma heating and current drive in fusion devices. Thus, it is  important to be able to simulate these waves with fidelity, which is where QC could, in principle, make a difference. If quantum modeling could be made significantly faster than classical one, that would, for example, help increase the spatial resolution of RF-wave simulations. Then, one would be able to resolve high-order cyclotron resonances and also more robustly calculate mode conversion \cite{book:tracy} when the emergence of electrostatic oscillations with small wavelengths makes fine grids necessary. This could be useful for accurate modeling of waves in the electron-cyclotron, lower-hybrid, and ion-cyclotron frequency ranges \cite{book:stix}. 

In \Sec{sec:cold}, we show that a broad class of linear RF plasma waves can be mapped to \Eq{eq:schr} with a sparse Hamiltonian. Specifically, they include all waves in cold collisionless static plasmas, which can be arbitrarily inhomogeneous. In \Sec{sec:linearnonherm}, we consider general fluid waves in plasma, which can exhibit instabilities or irreversible dissipation. Such waves are governed by pseudo-Hermitian or non-Hermitian sparse Hamiltonians, and we shall discuss possible approaches to mapping those on a quantum computer. (We shall also return to such waves in \Sec{sec:eigen} in the context of the eigenvalue problem.) In \Sec{sec:kinetic}, we consider kinetic waves. Depending on a problem, the corresponding Hamiltonians can also be Hermitian; however, unlike  for fluid waves, they are not sparse. Quantum simulations of such systems are less efficient, but we shall briefly outline some possibilities to deal with this issue.

%--------------------------------------
\subsection{Waves in cold collisionless plasmas: sparse Hermitian Hamiltonians}
\label{sec:cold}

%......................................
\subsubsection{Introduction}
\label{sec:rfintro}

It is well known that the equations governing electromagnetic waves in vacuum allow the Schr\"odinger representation \eq{eq:schr} for the ``photon wave function'', which is generally six-dimensional \cite{foot:photon}. Such waves can be modeled using QHS, for which a concrete algorithm has been recently proposed \cite{ref:costa19, ref:suau21}. The Schr\"odinger representation is also known for waves in inhomogeneous media described by real nondispersive dielectric permittivity and magnetic permeability \cite{my:qdiel}. For stable waves in nondissipative media with arbitrary dispersion, the Schr\"odinger representation has been proven to exist too \cite{my:wkin}. It can be found from general principles in the small-wavelength limit, for example, in the geometrical-optics and quasioptical approximations \cite{my:quasiop1, my:quasiop2, my:quasiop3}. However, deriving the actual Schr\"odinger representations for exact, or ``full-wave'', linear plasma-wave problems requires a detailed consideration of plasma dynamics. 

%......................................
\subsubsection{Basic equations}
\label{sec:basiceq}

There is at least one plasma model within which an exact Schr\"odinger representation of full-wave dynamics can be formulated explicitly and leads to sparse Hermitian Hamiltonians. This is the linearized model of cold collisionless static plasma, which is often sufficient for RF-wave modeling in practical applications (up to dissipation, which is discussed in \Sec{sec:linearnonherm}). Let us consider this model in detail. Suppose that plasma is formed by some~$\mc{N}$ species with charges~$e_s$, masses~$m_s$, and unperturbed densities $n_{0s} = n_{0s}(\vec{x})$, where~$\vec{x}$ is the spatial coordinate. The linearized equation for the fluid velocity~$\vec{v}_s$ of each species in a wave with electric field~$\vec{E}$~is
\begin{gather}\label{eq:vs}
\pd_t \vec{v}_s = (e_s/m_s)\,\vec{E} + \vec{v}_s \times \vec{\Omega}_s,
\end{gather}
where $\vec{\Omega}_s \doteq e_s \vec{B}_0(\vec{x})/(m_s c)$ is the $s$th-species gyrofrequency (the symbol $\doteq$ denotes definitions), $\vec{B}_0$ is the dc magnetic field, and $c$ is the speed of light. Consider a rescaled velocity $\vec{\zeta}_s \doteq \vec{v}_s(4\pi n_{0s} m_s)^{1/2}$, which has the same units as $\vec{E}$. Then, \Eq{eq:vs} becomes
\begin{gather}\label{eq:zetas}
\pd_t \vec{\zeta}_s = \omega_{ps}\,\vec{E} + \vec{\zeta}_s \times \vec{\Omega}_s,
\end{gather}
where $\omega_{ps} \doteq \smash{e_s (4\pi n_{0s}/m_s)^{1/2}}$ is the signed plasma frequency of species $s$. (This representation is also used in \Refs{ref:friedland88, my:covar, phd:ruiz17} for related calculations.) Let us complement this equation with Ampere's law and Faraday's laws,
\begin{gather}
\textstyle
\pd_t \vec{E} = - \sum_s \omega_{ps}\vec{\zeta}_s + c \del \times \vec{B},
\\
\pd_t \vec{B} = - c \del \times \vec{E},
\label{eq:farad}
\end{gather}
where $\vec{B}$ is the wave magnetic field. Using the Hermitian matrices
\begin{widetext}
\begin{gather}
\alpha_x = 
 \left(
 \begin{array}{rrr}
  0 & 0 & 0\\[0pt]
	 0 & 0 & -\ii\\[0pt]
	 0 & \ii & 0
 \end{array}
 \right), 
\quad
\alpha_y = 
 \left(
	\begin{array}{rrr}
	 0 & 0 & \ii\\[0pt]
	 0 & 0 & 0\\[0pt]
	-\ii & 0 & 0
	\end{array}
 \right), 
\quad
 \alpha_z = 
 \left(
	 \begin{array}{rrr}
	 0 & -\ii & 0\\[0pt]
		\ii & 0 & 0\\[0pt]
		0 & 0 & 0
	 \end{array}
	\right),
\end{gather}
one can also express the vector products through $\vec{\alpha} \doteq \{\alpha_x, \alpha_y, \alpha_z\}$. [Note that $\alpha_a$ are related to the Gell--Mann matrices, which serve as infinitesimal generators of SU$(3)$.] Specifically, for any three-component column vectors $\vec{\mc{A}}$ and $\vec{\mc{B}}$, one has $\vec{\mc{A}} \times \vec{\mc{B}} = - \ii(\vec{\alpha} \cdot \vec{\mc{A}}) \vec{\mc{B}}$, as can be verified by direct calculation. Then, \Eqs{eq:zetas}--\eq{eq:farad} can be written~as
\begin{gather}
\ii \pd_t \vec{\zeta}_s = \ii\omega_{ps}\,\vec{E} - (\vec{\alpha} \cdot \vec{\Omega}_s) \vec{\zeta}_s,
\\
\textstyle
\ii \pd_t \vec{E} = - \ii\sum_s\omega_{ps}\vec{\zeta}_s + \ii c (\vec{\alpha} \cdot \boper{k}) \vec{B},
\\
\ii \pd_t \vec{B} = - \ii c (\vec{\alpha} \cdot \boper{k}) \vec{E},
\end{gather}
where we have introduced the wavevector operator $\boper{k} \doteq - \ii\del$. These equations can be represented as a $3(\mc{N} + 2)$-dimensional vector equation of the form \eq{eq:schr} with $\psi = (8\pi)^{-1/2}\{\vec{\zeta}_1, \vec{\zeta}_2, \ldots, \vec{\zeta}_\mc{N}, \vec{E}, \vec{B}\}$ and a time-independent Hermitian Hamiltonian
\begin{gather}\label{eq:coldH}
\oper{H} = 
\left(
\begin{array}{cccccc}
- \vec{\alpha} \cdot \vec{\Omega}_1(\vec{x}) & 0 & \ldots & 0 & \ii\omega_{p1}(\vec{x}) & 0\\
0 & - \vec{\alpha} \cdot \vec{\Omega}_2(\vec{x}) & \ldots & 0 & \ii\omega_{p2}(\vec{x}) & 0\\
\vdots & \vdots & \ddots & \vdots & \vdots & \vdots \\
0 & 0 & \ldots & - \vec{\alpha} \cdot \vec{\Omega}_\mc{N}(\vec{x}) & \ii\omega_{p\mc{N}}(\vec{x}) & 0\\
- \ii\omega_{p1}(\vec{x}) & - \ii\omega_{p2}(\vec{x}) & \ldots & - \ii\omega_{p\mc{N}}(\vec{x}) & 0 & \ii c \vec{\alpha} \cdot \boper{k}\\
0 & 0 & \ldots & 0 & -\ii c \vec{\alpha} \cdot \boper{k}& 0
\end{array}
\right).
\end{gather}
This $\oper{H}$ is linear in $\boper{k}$, so it is naturally represented by a sparse matrix when mapped to a grid. Thus, efficient QHS of collisionless cold-plasma waves are, in principle, possible using already existing algorithms. Details, including specific algorithms and possible issues with the initial-state preparation, can be found in \Refs{ref:costa19, ref:gourdeau17}, where QHS for similar Hamiltonians have been recently discussed.

%......................................
\subsubsection{Relevant measurements} 
\label{sec:coldmeas}

For waves governed by Hamiltonians \eq{eq:coldH}, the output data of interest can be, say, the energy
\begin{gather}
\mcc{E} \doteq \int_V \Bigg(
\sum_s \frac{n_{0s} m_s v_s^2}{2} 
+ \frac{E^2}{8\pi}
+ \frac{B^2}{8\pi}
\Bigg)\,\dd\vec{x}
\end{gather}
within some finite volume~$V$, which can be expressed as $\mcc{E} = \int_V \psi^\dag(\vec{x}) \psi(\vec{x})\,\dd\vec{x}$. Let us introduce the window operator $\oper{\msf{W}} = \msf{W}(\vec{x})$ such that its coordinate representation is a window function $\msf{W}$ defined via $\msf{W}(\vec{x} \in V) = 1$ and $\msf{W}(\vec{x} \notin V) = 0$. Then, we can express the energy as $\mcc{E} = \int \psi^\dag(\vec{x}) \oper{\msf{W}} \psi(\vec{x})\,\dd\vec{x}$, where the integral is extended to the whole space. Hence, $\mcc{E}$ is the expectation value of $\oper{\msf{W}}$,
\begin{gather}
\mcc{E} = \braket{\psi|\oper{\msf{W}}|\psi},
\end{gather}
so it can be naturally extracted as an outcome of a quantum simulation. The local energy \textit{density} can be extracted as $\mcc{E}/V$ at $V \to 0$. Other quantities bilinear in $\psi$ can be extracted similarly too, by replacing $\oper{\msf{W}}$ with the appropriate operators.

%--------------------------------------
\subsection{General fluid waves: pseudo-Hermitian and non-Hermitian sparse Hamiltonians}
\label{sec:linearnonherm}

%......................................
\subsubsection{Introduction}

If fluid plasma has inhomogeneous density \textit{and} finite temperature (or average flow velocity), then it has free energy \cite{ref:gardner63, ref:bernstein58, my:restack, ref:helander17} that can drive linear instabilities. Although the corresponding dynamics remains Hamiltonian in the general sense of the word, the ``quantumlike'' Hamiltonian that enters the corresponding Schr\"odinger equation ceases to be Hermitian and becomes \textit{pseudo}-Hermitian instead \cite{ref:larsson91, ref:brizard94, ref:mostafazadeh02}. This means that plasma dynamics is governed by
\begin{gather}\label{eq:pseudo1}
\ii \pd_t \psi = \oper{H} \psi, \quad \oper{H}^\dag \oper{\eta} = \oper{\eta}\oper{H},
\end{gather}
where $\oper{\eta}$ is some time-independent Hermitian operator. (For example, see \Ref{ref:brizard92} for the absence of Hermiticity in linearized MHD and also \Refs{my:wkeadv, my:shear, ref:qin19} for the absence of Hermiticity in hydrodynamical perturbations in sheared flows.) Unless $\oper{\eta}$ is positively defined, there is no variable transformation that maps \Eq{eq:pseudo1} to \Eq{eq:schr} and \Eq{eq:pseudo1} cannot be solved directly using QHS.

The same conclusion applies if plasma is collisional. For example, consider cold electron-ion plasma, possibly with immobile neutrals in the background. Then, the electron and ion velocities satisfy
\begin{gather}
\pd_t \vec{v}_e = (e_e/m_e)\,\vec{E} + \vec{v}_e \times \vec{\Omega}_e 
- \nu_{en}\vec{v}_e - \nu_{ei}(\vec{v}_e - \vec{v}_i),
\\
\pd_t \vec{v}_i = (e_i/m_i)\,\vec{E} + \vec{v}_i \times \vec{\Omega}_i 
- \nu_{in}\vec{v}_i - \nu_{ie}(\vec{v}_i - \vec{v}_e),
\end{gather}
where $\nu_{en}$ and $\nu_{in}$ are the electron and ion rates of collisions with neutrals respectively, $\nu_{ei}$ is the electron--ion collision rate, $\nu_{ie} = (Z m_e/m_i)\nu_{ei}$ is the ion--electron collision rate, and $Z$ is the ion charge state. The Hamiltonian that governs $\psi \doteq (8\pi)^{-1/2}\{
\vec{\zeta}_e, \vec{\zeta}_i, \vec{E}, \vec{B}\}$~is
\begin{gather}\label{eq:Hu}
\oper{H} = 
\left(
\begin{array}{cccc}
- \vec{\alpha} \cdot \vec{\Omega}_e - \ii (\nu_{en}+\nu_{ei}) & \ii \varsigma & \ii\omega_{pe} & 0\\
\ii\varsigma & - \vec{\alpha} \cdot \vec{\Omega}_i - \ii (\nu_{in}+\nu_{ie}) & \ii\omega_{pi} & 0\\
- \ii\omega_{pe} & - \ii\omega_{pi} & 0 & \ii c \vec{\alpha} \cdot \boper{k}\\
0 & 0 & -\ii c \vec{\alpha} \cdot \boper{k}& 0
\end{array}
\right),
\end{gather}
\end{widetext}
where $\varsigma \doteq \nu_{ei} \sqrt{Z m_e/m_i}$ and $\vec{x}$-dependence of the coefficients is allowed, like in \Eq{eq:coldH}. (The remaining notation is the same as in \Sec{sec:cold}.)

For non-Hermitian systems like those governed by \Eqs{eq:pseudo1} and \eq{eq:Hu}, some authors proposed methods close to QHS \cite{ref:motta19, ref:candia15}, but those methods are unlikely to suit plasma simulations. Instead, we propose to follow the idea from \Ref{ref:berry14}, which is as follows. 

%......................................
\subsubsection{Initial-value problem} 
\label{sec:initv}

Let us consider time as one of the coordinate variables and introduce the corresponding ``momentum'' (frequency, or energy) operator $\oper{\omega} \doteq \ii \pd_t$. Let us also introduce $\oper{\mc{H}} \doteq \oper{\omega} - \oper{H}$. Then, one can rewrite \Eq{eq:schr} as $\oper{\mc{H}}\psi = \xi$, with $\xi(t, \vec{x}) \doteq \ii\psi_0(\vec{x})\delta(t)$, where $\delta$ is the Dirac delta function. On a time grid $t = \lbrace t_0, t_1, \ldots \rbrace$, where $t_0 = 0$, this becomes $\oper{\mc{H}}\psi = \Xi$, where $\psi \doteq \lbrace \psi(t_0, \vec{x}), \psi(t_1, \vec{x}), \ldots \rbrace$,
\begin{gather}\label{eq:Xin}
\Xi_n(\vec{x}) = \ii\psi_0(\vec{x})\delta_{n,0},
\end{gather} 
and $\delta_{a,b}$ is the Kronecker symbol. This equation can be represented as
\begin{gather}\label{eq:AXY}
\oper{\msf{A}} \msf{X} = \msf{Y},
\end{gather}
where $\oper{\msf{A}}$ is a Hermitian operator. Specifically,
\begin{gather}\label{eq:AAA}
\oper{\msf{A}} \doteq
\left(
\begin{array}{cc}
0 & \oper{\mc{H}}\\
\oper{\mc{H}}^\dag & 0
\end{array}
\right),
\quad
\msf{X} \doteq 
\left(
\begin{array}{c}
0 \\ \psi
\end{array}
\right),
\quad
\msf{Y} \doteq 
\left(
\begin{array}{c}
\Xi \\ 0
\end{array}
\right)
.
\end{gather}
(Dissipation and instabilities are captured within this approach in the structure of the eigenvectors of $\oper{\msf{A}}$. In a way, these eigenvectors can be understood as ``surface modes'' bounded on the time axis to the initial and finite moments of time.) If \Eq{eq:AXY} is also discretized in space, then the nontrivial part of $\msf{Y}$ has dimension much less than $\dim\msf{X}$ due to the Kronecker symbol in \Eq{eq:Xin}, so the right-hand side of \Eq{eq:AXY} on a grid can be prepared efficiently. Then in principle, this equation can be solved efficiently using the known Harrow--Hassidim--Lloyd (HHL) or other quantum algorithms \cite{ref:harrow09, ref:childs17}. 

Naturally, those algorithms are not a magic wand; for example, they are efficient only for sparse matrices and require that the condition number scales well. A discussion of these problems is beyond the scope of our paper, but see \Refs{ref:clader13, ref:scherer17, ref:montanaro16b}. Let us only point out one issue, which is less technical. The eigenvalues $\lambda$ of the Hermitian operator $\oper{A}$ can be related to those of the original non-Hermitian operator $\oper{\mc{H}}$. By definition,
\begin{gather}\label{eq:detl}
\det \left(
\begin{array}{cc}
- \lambda \unitoper & \oper{\mc{H}}\\
\oper{\mc{H}}^\dag & - \lambda \unitoper
\end{array}
\right) = 0, 
\end{gather}
where $\unitoper$ is a unit operator. Using Schur's determinant identity, one can rewrite \Eq{eq:detl} as follows:
\begin{multline}
0 = \det(- \lambda \unitoper)\,\det[- \lambda \unitoper - \oper{\mc{H}}(- \lambda^{-1} \unitoper)\oper{\mc{H}}^\dag] 
\\ = \det(\lambda^2 \unitoper - \oper{\mc{H}}\oper{\mc{H}}^\dag).
\end{multline}
Hence, $\lambda$ can be found as the (real) eigenvalues of $\smash{\pm (\oper{\mc{H}}\oper{\mc{H}}^\dag)^{1/2}}$. This shows that $\lambda$ may not depend analytically on the parameters of a problem even when $\oper{H}$ does. (In the special case when $\oper{\mc{H}}$ is Hermitian, $\lambda$ are simply the eigenvalues of $\smash{\pm \oper{\mc{H}}}$; then, they are analytic if $\oper{H}$ is analytic~\cite{ref:mengi14}.) To what extent this affects of robustness of the whole scheme is yet to be determined.

%......................................
\subsubsection{Boundary-value problem} 
\label{sec:boundary}

RF-wave simulations in plasma physics are typically concerned with stationary waves, in which case the frequency $\omega$ is constant and $\psi$ prescribed on some boundary, say, an antenna. Then, instead of solving an initial-value problem, one can solve a boundary-value problem, which is even simpler. In this case, $\oper{\omega} = \omega$ is a constant and \Eq{eq:schr} can be expressed as $\oper{\msf{H}} \psi = 0$, where $\oper{\msf{H}} \doteq \omega - \oper{H}$. Let us assume the decomposition
\begin{gather}
\oper{\msf{H}} \doteq
\left(
\begin{array}{cc}
\oper{\msf{H}}_{aa} & \oper{\msf{H}}_{ab}\\
\oper{\msf{H}}_{ba} & \oper{\msf{H}}_{bb}
\end{array}
\right),
\quad
\psi =
\left(
\begin{array}{c}
a \\ b
\end{array}
\right),
\end{gather}
where $b$ is the part of $\psi$ that belongs to the antenna. Then, $a$ is governed by 
\begin{gather}\label{eq:Hbound}
\oper{\mc{H}}a = \Xi, \quad \oper{\mc{H}} \doteq \oper{\msf{H}}_{aa}, \quad \Xi \doteq - \oper{\msf{H}}_{ab}b. 
\end{gather}
Note that $\dim b$ is proportional, with a small coefficient, to the number of cells representing the plasma surface, while $\dim a$ is roughly the number of cells representing the plasma volume, $\dim a \gg \dim b$. This means that $b$ can be prepared efficiently, and thus so can $\Xi$. (Remember that $\oper{\msf{H}}_{ab}$ is sparse.) Furthermore, \Eq{eq:Hbound} has the same form as the one in the initial-value problem. Thus, in principle, this equation can be solved efficiently using the same method as in \Sec{sec:initv}, with the same reservations.

%......................................
\subsubsection{Relevant measurements} 

For dissipative linear waves, the result sought in simulations is typically the power $P_{\rm abs} = \int_V \mc{P}_{\rm abs}\,\dd\vec{x}$ dissipated in some finite volume~$V$. (If dissipation is mainly resonant, it can be assumed well localized in space, so~$V$ can be small compared to the simulation box.) Most generally, $\mc{P}_{\rm abs}$ can be related to the anti-Hermitian part of the Hamiltonian $\oper{H}$; for example, see \Ref{my:zonal}. However, it is often enough to calculate this power within the geometrical-optics approximation \cite{book:stix},
\begin{gather}\label{eq:Pabs1}
\mc{P}_{\rm abs} = \frac{\omega}{4\pi}\,
\favr{\vec{E}^\intercal\vec{\epsilon}_A(t, \vec{x}, \omega, \vec{k}) \vec{E}}_t.
\end{gather}
Here, $\vec{E}^\intercal = \vec{E}^\dag$ is the transposed (real) electric-field vector, $\vec{\epsilon}_A = \vec{\epsilon}_A^\dag \doteq (\vec{\epsilon} - \vec{\epsilon}^\dag)/2\ii$, $\vec{\epsilon}$ is the dielectric tensor that slowly depends on $t$ and $\vec{x}$, $\omega$ and $\vec{k}$ are the local frequency and the local wavevector, and $\favr{\ldots }_t$ denotes time averaging over the wave period. Within the geometrical-optics approximation, one can replace \Eq{eq:Pabs1} with $\mc{P}_{\rm abs} = \left\langle\vec{E}^\intercal \oper{\msf{P}} \vec{E} \right\rangle_t$,
\begin{gather}
\oper{\msf{P}} \doteq \frac{1}{8\pi}
\left\{
\oper{\omega}\vec{\epsilon}_A(t, \vec{x}, \oper{\omega}, \boper{k})
+
\big[\oper{\omega}\vec{\epsilon}_A(t, \vec{x}, \oper{\omega}, \boper{k})\big]^\dag
\right\}.
\end{gather}
Then, the dissipated power can be expressed as the following expectation value:
\begin{gather}
P_{\rm abs} \propto \braket{\vec{E}| \oper{\msf{P}}\oper{\msf{W}} + \oper{\msf{W}} \oper{\msf{P}}| \vec{E}},
\end{gather}
where $\oper{\msf{W}}$ is a window operator. For a boundary problem, $\oper{\msf{W}}$ is the same as in \Sec{sec:coldmeas} (and $\oper{\omega} = \omega$ is a real constant). For an initial-value problem, the window function must be defined in spacetime, and the length of that along the time axis must be much larger than the characteristic temporal period $2\pi/\omega$.

%--------------------------------------
\subsection{Kinetic waves}
\label{sec:kinetic}

Now, let us discuss the possibility of quantum simulations of kinetic waves. For simplicity,\footnote{A quantumlike formulation of the \textit{general} linearized Vlasov--Maxwell system is also possible \cite{ref:larsson91} but requires a more complicated definition of the state function, so we do not consider the general case in this preliminary study.} let us limit our discussion to the collisionless kinetic model where the background plasma is homogeneous and isotropic. For spatially monochromatic fields in Maxwellian plasma, this model was previously discussed in \Ref{ref:engel19}, but here, we present it in a somewhat more general form. In particular, we do not restrict the field profile, and our general approach can be readily extended to inhomogeneous nonisotropic plasmas with flows.

Let us assume the distribution function of species $s$ in the form $f_s(t, \vec{x}, \vec{p}) = f_{0s}(\vec{p}) + \tilde{f}_s(t, \vec{x}, \vec{p})$. Here, $f_{0s}$ is the background distribution and $\tilde{f}_s \ll f_{0s}$ is a small perturbation that satisfies the linearized Vlasov equation
\begin{multline}\label{eq:tf}
\pd_t \tilde{f}_s + \vec{v}_s \cdot \del \tilde{f}_s 
+ e_s(\vec{v}_s \times \vec{B}_0/c)\cdot \pd_\vec{p} \tilde{f}_s
\\= - e_s(\vec{E} + \vec{v}_s \times \vec{B}/c)\cdot\pd_\vec{p} f_{0s}.
\end{multline}
Here, $\vec{v}_s \doteq \vec{p}/(\gamma_s m_s)$ and $\gamma_s \doteq (1 + p^2/m_s^2c^2)^{1/2}$ is the Lorentz factor. (We retain relativistic effects because keeping them does not significantly complicate our model.) Let us assume that the background distribution is isotropic, which we express as follows:
\begin{gather}
 f_{0s}(\vec{p}) = F_s(\mc{E}_s(\vec{p})/T_s).
\end{gather}
Here, $\mc{E}_s =\gamma_s m_s c^2$ is the energy, $T_s > 0$ is some effective temperature or \textit{the} temperature, if the distribution is Maxwellian. Then, $\pd_\vec{p} f_{0s} = -\vec{v}_s F_s'/(m_s T_s)$, so $\vec{v}_s \times \vec{B} \cdot \pd_\vec{p} f_{0s} = 0$, and \Eq{eq:tf} becomes
\begin{gather}\label{eq:ttf}
\ii \pd_t \tilde{f}_s = \oper{h}_s \tilde{f}_s 
- \ii e_s \vec{E} \cdot \vec{v}_s F_s'/T_s.
\end{gather}
Here, $\oper{h}_s$ is an operator that is Hermitian on the phase space $\msf{z} \doteq (\vec{x}, \vec{p})$ under the Euclidean metric; specifically,
\begin{align}
\oper{h}_s \tilde{f}_s 
& \doteq \vec{v}_s \cdot (-\ii \del) \tilde{f}_s
+ (\vec{v}_s \times \vec{B}_0/c)\cdot(-\ii\pd_\vec{p}) \tilde{f}_s
\\ 
& =
 -\ii \del \cdot (\vec{v}_s \tilde{f}_s)
-\ii\pd_\vec{p} \cdot[ (\vec{v}_s \times \vec{B}_0/c) \tilde{f}_s].
\end{align}

In order to make \Eq{eq:ttf} manifestly conservative in conjunction with Ampere's law, consider a rescaled distribution $g_s(t, \msf{z}) \doteq \tilde{f}_s(t, \msf{z})/[\ii r_s(\mc{E}_s)]$ with $r_s = \sqrt{|F_s'|/4\pi T_s}$. Then, one obtains
\begin{gather}\label{eq:Eg1}
\textstyle
\ii\pd_t \vec{E} = \ii c (\vec{\alpha} \cdot \boper{k}) \vec{B}
+ \sum_s \int \dd\vec{p}\, R_s \vec{v}_s g_s,
\\
\textstyle
\ii\pd_t g_s = \oper{h}_s g_s + \sigma_s R_s \vec{E} \cdot \vec{v}_s,
\label{eq:Eg2}
\end{gather}
where $R_s \doteq e_s\sqrt{4\pi |F_s'|/T_s}$ and $\sigma_s \doteq \textrm{sign}\,(-F_s')$. Finally, let us discretize the momentum space, so $\int \dd\vec{p} \mapsto \sum_{\vec{p}} \smash{(\Delta p)^3}$, and rescale $g_s \mapsto \smash{(\Delta p)^{-3/2}} g_s$ and $R_s \mapsto \smash{(\Delta p)^{-3/2}} R_s$. Then, the resulting model is as follows:
\begin{gather}
\textstyle
\ii\pd_t \vec{E} = \ii c (\vec{\alpha} \cdot \boper{k}) \vec{B}
+ \sum_{s,\vec{p}} R_s \vec{v}_s g_s,
\label{eq:EBg1}
\\
\ii\pd_t \vec{B} = -\ii c (\vec{\alpha} \cdot \boper{k}) \vec{E},
\label{eq:EBg2}
\\
\textstyle
\ii\pd_t g_s = \oper{h}_s g_s + \sigma_s R_s \vec{E} \cdot \vec{v}_s,
\label{eq:EBg3}
\end{gather}
where we have included Faraday's law for completeness.

Equations \eq{eq:EBg1} form a Schr\"odinger-type equation for the vector field $\psi = (8\pi)^{-1/2}\{\vec{E}, \vec{B}, \vec{g}\}$, where each element of the vector $\vec{g}$ is a field in the $\vec{x}$ space, $g_s(t, \vec{x}, \vec{p})$, in which $s$ and $\vec{p}$ are fixed parameters. Like in the previous sections, relevant quantities of interest in this case are bilinear functionals of $\psi$, or the expectation values of (spatial or phase-space) window operators and other linear operators. Also, the corresponding Hamiltonian can be symbolically expressed as follows:
\begin{gather}\label{eq:Hkin}
\oper{H} = 
\left(
\begin{array}{ccc}
0 & \ii c (\vec{\alpha} \cdot \boper{k}) & R \vec{v} \\
-\ii c (\vec{\alpha} \cdot \boper{k}) & 0 & 0\\
\sigma R \vec{v}  & 0 & \oper{h} 
\end{array}
\right).
\end{gather}

If some $F_s$ are nonmonotonic ($\sigma_s \ne 1$), meaning that the plasma has free energy, this Hamiltonian is pseudo-Hermitian and can support instabilities, as expected. Otherwise ($\sigma_s = 1$), $\oper{H}$ is Hermitian and the corresponding plasma dynamics can, in principle, be modeled using QHS. However, note that $\oper{H}$ is not sparse, so QHS are less efficient for kinetic simulations than for cold-wave simulations. This problem was addressed in \Ref{ref:engel19}. (The model from \Ref{ref:engel19} is obtained from ours as a special case by assuming Maxwellian plasma and $\boper{k} = \vec{k}$.) There, the authors adopted the approach from \Refs{ref:low17, ref:low19, tex:low17, ref:low16}, which formally allows efficient QHS with arbitrary non-sparse Hamiltonians. The recent study \cite{ref:childs18} indicates that this approach may be challenging beyond toy problems\footnote{In this approach, the evolution operator $\exp(-\ii \oper{H}t)$ is represented through a series of rotations whose angles are found numerically. The authors of \Ref{ref:childs18} ``were unable to compute [those angles] explicitly except in very small instances''.}, so its practicality remains to be determined; but other approaches may also be possible. Note that the parts of the distribution function $\tilde{f}_s$ corresponding to different velocity elements interact with each other only through the collective electric field rather than directly. This means that \textit{the graph of the $\oper{H}$ is a star} (or more precisely, star with loops, due to the diagonal terms). Such special structure potentially allows for efficient QHS \cite{ref:childs09, ref:loke12}, although explicit algorithms for modeling kinetic plasma waves are yet to be developed.

%%%%%%%%%%%%%%%%%%%%%%%%%%%%%%%%%%%%%%%
\section{Nonlinear dynamics}
\label{sec:nonlin}

%--------------------------------------
\subsection{Preliminary considerations}
\label{sec:prelim}

Suppose a generic ODE
\begin{gather}\label{eq:uG}
\dot{u} = g(t, u).
\end{gather}
Here, the dot denotes a derivative with respect to time $t$, $u \equiv u(t, u_0)$ is some vector $\lbrace u^1, u^2, \ldots, u^{d_u}\rbrace$, $u_0$ is a given initial value serving as a parameter, and $g \equiv \lbrace g^1, g^2, \ldots, g^{d_u} \rbrace$ is a vector function that may be nonlinear. (The upper indices denote the vector components and must not be confused with power indices.)

In the ``standard'' quantum algorithm for nonlinear ODEs proposed in \Ref{tex:leyton08}, $u$ is encoded in the amplitude of the state function such that $\psi \,\propto\, u$. Suppose a simple nonlinearity, say, $g\, \propto\, u^2$. Then, \Eqs{eq:uG} can be solved on a quantum computer iteratively if there is a subroutine that can generate a state
\begin{gather}\label{nl:2}
\textstyle
\ket{\psi'}= \sum_{ijk} A_{ijk} \psi_j \psi_k \ket{i}
\end{gather}
from a given state $\ket{\psi}=\sum_i \psi_i\ket{i}$. The nonlinear transformation $\ket{\psi} \mapsto \ket{\psi'}$ cannot be produced with a single copy of $\ket{\psi}$ due to the linear nature of quantum mechanics and the so-called no-cloning theorem \cite{book:nielsen}. Still, it can be produced with a unitary transformation $\exp(-\ii\epsilon H)$ (with $\epsilon \ll 1$) if one has two copies of $\ket{\psi}$ and an additional ancilla qubit $P$ in the state initialized to $\ket{0}_P$,
\begin{multline}
\ket{\psi}\ket{\psi}\ket{0}_P\mapsto\exp(-\ii\epsilon\oper{H})\ket{\psi}\ket{\psi}\ket{0}_P 
\\ \approx \ket{\psi}\ket{\psi}\ket{0}_P + \epsilon \oper{A}\ket{\psi}\ket{\psi}\ket{1}_P. \label{nl:3}
\end{multline}
Here, the two-state non-Hermitian operator $\oper{A}$ is given~by
\begin{gather}
\textstyle
\oper{A} =\sum_{ijk} A_{ijk}\ket{i,0}\bra{j,k}  \label{nl:4}
\end{gather}
and acts as follows:
\begin{gather}
\textstyle
\oper{A} \ket{\psi}\ket{\psi}=
\left(\sum_{ijk} A_{ijk}\psi_j \psi_k \ket{i}\right)\ket{0}.  \label{nl:5}
\end{gather}
Also, the Hermitian Hamiltonian $\oper{H}$ is constructed to implement ``von~Neumann measurement operation'', which entangles the desired result with the ancilla qubit $P$ \cite{tex:leyton08},
\begin{gather}
\oper{H}=-\ii \oper{A} \otimes \ket{1}_P \bra{0}_P
+ \ii \oper{A}^\dag\otimes\ket{0}_P\bra{1}_P. \label{nl:6}
\end{gather}
Then, measuring the ancilla qubit $P$ in the resulting state~\eq{nl:3} and post-selecting the results with $P$ in the state $\ket{1}_P$ results in the desired state $\ket{\psi'}$ with probability $\sim \epsilon^2$. Alternatively one can use the amplitude-amplification algorithm \cite{tex:brassard00} that requires $\sim 1/\epsilon$ operations to increase the amplitude of the the state with $\ket{1}_P$ to $\sim 1/2$. In either case, at least two copies of the state $\ket{\psi}$ are required at every iteration step, which are then replaced by one copy of $\ket{\psi'}$ by the algorithm. This means that the number of copies of the initial state scales exponentially  with the number of steps. Furthermore, this method is effectively restricted to $g$ that are low-order polynomials of $u$. Hence, it is unlikely to be suitable for practical ODE solvers.\footnote{That said, this algorithm is advantageous in that the number of qubits it requires scales logarithmically with the number of degrees of freedom. We shall return to this in \Sec{sec:subdiscussion}.}

The alternative is to convert a nonlinear problem \eq{eq:uG} into a linear one. Although some nonlinear equations allow \textit{ad~hoc} variable transformations that make them linear, such special cases are of limited interest in practice. A more reliable approach is to extend the configuration space by introducing sufficiently many auxiliary degrees of freedom. Sometimes, adding a single degree of freedom is already enough (\App{app:llg}), but here we shall focus on methods that are more universal. In \Sec{sec:ham}, we consider the case of classical Hamiltonian dynamics, and the most general case is considered in \Sec{sec:general}. Yet another, variational, approach to nonlinear simulations, which is based on hybrid quantum--classical computing, will be discussed in \Sec{sec:varia}.

%--------------------------------------
\subsection{Classical Hamiltonian systems}
\label{sec:ham}

Classical Hamiltonian systems can always be made linear via quantization. For example, suppose that \Eqs{eq:uG} have the form
\begin{gather}\label{eq:zH}
\dot{x}^a = \pd_{p_a} \mc{H},
\quad
\dot{p}_a = -\pd_{x^a} \mc{H},
\end{gather}
where $\mc{H} = \mc{H}(t, x, p)$ is some scalar function known as the Hamiltonian. This system can be mapped to a linear quantum system
\begin{gather}\label{eq:heff}
\ii \heff \pd_t \psi = \oper{\mc{H}} \psi.
\end{gather}
Here, $\psi$ is some complex scalar field and $\heff$ is a fake Planck constant that is introduced arbitrarily such that it be small enough but not necessarily equal (or even comparable) to the true Planck constant. The operator $\oper{\mc{H}}$ can be obtained from $\mc{H}$ by, say, taking the Weyl transform of the latter.\footnote{For example, see \Ref{book:tracy} or the supplemental material in \Ref{my:quasiop1}.} A procedure that is less pleasing aesthetically but still sufficient is to replace $x^a$ with the coordinate operator $\oper{x}^a$ (assuming the coordinate space is Euclidean), replace $p_a$ with the momentum operator $-\ii \heff \pd_{x^a}$, and then take the Hermitian part of the resulting operator. As long as the effective de Broglie wavelength associated with $\psi$ remains small compared to the characteristic scales of the problem, the dynamics generated by \Eq{eq:heff} will adequately reflect the dynamics of the original classical system, and the classical variables can be found as expectation values of $\psi$.

For example, let us consider $\mc{H}$ that is the Hamiltonian of a nonrelativistic classical particle interacting with electromagnetic field:
\begin{gather}
\mc{H}(t, \vec{x}, \vec{p}) = \frac{1}{2m}\left[
\vec{p}-\frac{e}{c}\,\vec{A}(t, \vec{x})
\right]^2 + e\varphi(t, \vec{x}).
\end{gather}
Here, $m$ and $e$ are the particle mass and charge, $\vec{A}$ is a vector potential, and $\varphi$ is a scalar potential. Then,
\begin{gather}
\oper{\mc{H}} = \frac{1}{2m}\left[
\oper{\vec{p}}-\frac{e}{c}\,\vec{A}(t, \oper{\vec{x}})
\right]^2 + e\varphi(t, \oper{\vec{x}}),
\end{gather}
which is Hermitian already as is (\ie hermitization is not needed in this case). Assuming the Madelung representation $\psi = \sqrt{n}\,\ee^{\ii\theta/\heff}$, where $n$ and $\theta$ are real, one obtains (see, \eg \Ref{my:qlagr})
\begin{gather}
\pd_t n + \del \cdot (n \vec{v}) = 0,
\label{eq:n}
\\
m (\pd_t + \vec{v} \cdot \del) \vec{v} 
= e (\vec{E} + \vec{v} \times \vec{B}/c)- \del Q,
\label{eq:v}
\end{gather}
where $\vec{v} \doteq (\nabla \theta - e\vec{A}/c)/m$ is the velocity, $\vec{E} \doteq - (1/c)\pd_t \vec{A} - \del \varphi$ and $\vec{B} \doteq \del \times \vec{A}$ are the electric and magnetic fields, and $Q$ is the Bohm potential, which is given by
\begin{gather}
Q = - \frac{\heff^2}{2m}\,\frac{\del^2 \sqrt{n}}{\sqrt{n}}.
\end{gather}
At small enough $\heff$, the Bohm potential is negligible (assuming that the characteristic spatial scale of $n$ is independent of $\heff$), so one obtains a semiclassical model, whose characteristics are exactly \Eqs{eq:zH}.

Notably, albeit not surprisingly, \Eqs{eq:n} and \eq{eq:v} are just the classical equations of cold charged fluid with density $n$ and velocity $\vec{v}$. In this sense, our approach allows solving not only discrete Hamilton's equations but nonlinear fluid equations as well. The only subtlety is that, by definition,
\begin{gather}\label{eq:irr}
\del \times (m\vec{v} + e\vec{A}/c) = \del \times \del \theta \equiv 0.
\end{gather}
(See also \Ref{ref:seliger68}, which elaborates on the related issue in the variational formulation of classical fluid mechanics.) If more general fluids need to be modeled, they can be represented as ensembles of fluids satisfying \Eq{eq:irr}; \ie multiple functions $\psi$ can be introduced. Also note that alternative approaches to quantum simulations of classical fluids within the Navier--Stokes model were recently discussed in \Refs{tex:budinski21, ref:gaitan20}.

%--------------------------------------
\subsection{General approach}
\label{sec:general}

Now, let us return to the general \Eq{eq:uG}. We shall assume that both $u$ and $g$ are real; otherwise, the real and imaginary parts of $u$ can be treated as independent components of a real vector that satisfies an equation of the form \eq{eq:uG}. Consider\footnote{A similar approach was also proposed in parallel in \Ref{ref:joseph20}. Since the first preprint of our paper was released, related ideas have also been proposed in \Refs{tex:engel20, tex:liu20, tex:lloyd20}.}
\begin{gather}\label{eq:Fdelta}
F(t, w) = \delta[w-u(t, u_0)]
\end{gather}
(as a reminder, $\delta$ is the Dirac delta function), which represents the probability distribution in space $w$ that corresponds to the solution $w = u (t,u_0)$ with specific~$u_0$. Then, one obtains
\begin{gather}
\pd_t F
=-\dot{u}^a [\pd_{w^a}\delta (w-u)] =-\pd_{w^a}[g^a(t,w)\delta (w-u)],
\nonumber
\end{gather}
where summation over repeated indices is assumed. This can be viewed as a linear continuity equation for $F$,
\begin{gather}
\pd_t F(t, w) + \nabla_w \cdot [g(t, w)F(t, w)] = 0.
\label{eq:Fcont}
\end{gather}
Next, let us introduce $\psi \doteq \sqrt{F(t, w)}$. [For simplicity, one can consider $F = \delta(w-u)$ as a sufficiently narrow Gaussian; then $\sqrt{F}$ is defined as usual \cite{ref:craven85}. For a general definition and for how to map such objects to a grid, see \App{app:delta}.] This function satisfies
\begin{gather}
\pd_t\psi 
=-\frac{1}{2}\,(\nabla_w \cdot g)\psi - g \cdot \nabla_w\psi.
\label{eq:psit}
\end{gather}
To rewrite this in a compact form, let us introduce the coordinate operators $\oper{w}^a$ on the the $w$ space and the corresponding momentum operators $\oper{\rho}_a$:
\begin{gather}
\oper{\rho}_a \doteq -\ii\pd_{w^a},
\quad
[\oper{w}^a, \oper{\rho}_b] = \ii \delta_b^a,
\end{gather}
where $[\cdot, \cdot]$ is a commutator. Then, $g$ can also be viewed as an operator, $\oper{g}^a\doteq g^a(t, \oper{w})$, which is Hermitian, because $g^a$ is real. Accordingly, the above equation for $\psi$ can be expressed as
\begin{gather}
\ii \pd_t\psi 
=\frac{1}{2}\,
[\oper{\rho}_a, \oper{g}^a]\psi + \oper{g}^a \oper{\rho}_a
\psi = \oper{H}\psi,
\label{eq:psit2}
\end{gather}
where $\oper{H}$ is a linear \textit{Hermitian} operator given by
\begin{gather}\label{eq:HL}
\oper{H}=\frac{1}{2}\,
(\oper{\rho}_a \oper{g}^a + \oper{g}^a \oper{\rho}_a).
\end{gather}

Equation \eq{eq:psit2} has the form of a geometrical-optics wave equation \cite{my:quasiop1}. It is also a Schr\"odinger equation with a sparse Hamiltonian, so it can be solved directly using QHS. Once the solution for $\psi$ has been obtained, the value of $u^a$ at any given $t$, which can be expressed as $u^a(t) = \int \delta (w-u(t, u_0))\,w^a \,\dd w$, is readily found as the expectation value of $\oper{w}^a$ on~$\psi$:
\begin{gather}\label{eq:psiw}
u^a =\int [\psi(t, w)]^2 w^a \dd w
\equiv \braket{\psi | \oper{w}^a | \psi},
\end{gather}
where we have used the fact that $\psi$ is real by definition.

Note that mapping the nonlinear problem \eq{eq:uG} to the linear problem \eq{eq:psit2} is exact. (Discretization errors occur when the equations are mapped on a grid, but they are not different from those in classical simulations of linear systems.) A disadvantage of this approach is that simulating the dynamics in the $w$ space is computationally expensive; it requires a grid whose number of cells scales as $\smash{N_w \sim n_u^{d_u}}$, where $n_u$ is the number of cells on the $u^a$ axis (assuming for simplicity that $n_u$ is the same for all~$a$). Since the required number of qubits scales logarithmically with $N_w$, it thereby scales linearly with $d_u$. This imposes limitations on how many degrees of freedom $d_u$ can be handled in practice. For example, this may not be a practical approach for solving partial differential equations, because they correspond to large $d_u$ when mapped on a grid. However, this approach is advantageous compared to the one described in \Sec{sec:prelim} in that its requirements on the computational resources do not grow exponentially with time and no intermediate measurements are involved. 

Also note that the same method can be used at no extra cost to model the evolution of $u$ averaged over any given initial distribution $f_0(u_0)$. The only difference in this case is that instead of \Eq{eq:Fdelta}, $F$ is defined as follows:
\begin{gather}\label{eq:F0}
F(t, w) = \int \delta[w - u(t, u_0)] f_0(u_0)\,\dd u_0.
\end{gather}
It may appear surprising that such linear superposition of solutions corresponding to different $u_0$ maps to a linear equation \eq{eq:psit2} even though $\psi \doteq \sqrt{F}$ depends on $F$ nonlinearly. But this is understood if one considers the problem on a grid. In this case, the continuous distribution $F$ splits into a sum of delta distributions, $F = \sum_n F_n$, and $F_n F_m \equiv 0$ for all $n \ne m$, because trajectories do not intersect. Thus, $\psi \doteq \sqrt{F}$ maps to the sum $\psi = \sum_n \psi_n$, where $\psi_n \doteq \sqrt{F_n}$ evolve independently, each with its own~$u_0$.

In case of Hamiltonian dynamics, when ${\nabla_w \cdot g} = 0$, \Eq{eq:Fcont} becomes the Liouville equation, with $w$ being the phase-space coordinate, and coincides with the Schr\"odinger equation for $\psi$. (The fact that the Liouville equation can be viewed as a Schr\"odinger equation has long been known; for example, see \Ref{my:wkin} and references therein.) In this case, the method described here can be viewed as a phase-space reformulation of the method described in \Sec{sec:ham}. Although the dimension of $u$ is twice as large as the dimension of $x$ ($d_u = 2d_x$), the de~Broglie wavelength does not need to be resolved, so both approaches require about the same number of cells in the corresponding $w$~spaces. In a given application, the first (\Sec{sec:ham}) or the second (\Sec{sec:general}) approach may be advantageous depending, for example, on a specific Hamiltonian.

%--------------------------------------
\subsection{Stochastic differential equations}

Another interesting class of problems is where the right-hand side of an ODE contains a stochastic term~$f^\alpha$:
\begin{gather}\label{eq:Ed0}
\dot{u}^\alpha=g^\alpha(t,u)+f^\alpha(t).
\end{gather}
Let us assume that $f^\alpha$ has Gaussian statistics with
\begin{gather}\label{eq:Ed1}
\favr{f^\alpha(t)} = 0,
\quad
\favr{f^\alpha(t) f^\beta(t')} = A^{\alpha\beta}\delta(t-t').
\end{gather}
Then, the corresponding equation \eq{eq:Fcont} for $F$ acquires an additional term: 
\begin{gather}
\pd_t F =-\pd_{w^\alpha}[g^\alpha(t,w)F]-\partial_{w^\alpha}\favr{f^\alpha\delta(w-u)}.
\label{eq:Ed2}
\end{gather}
Using the Novikov formula for a Gaussian noise \cite{ref:novikov65, book:mccomb},
\begin{gather}
\favr{f^\alpha(t) J[f]} = A^{\alpha\beta} \left\langle \frac{\delta J[f]}{\delta f^\beta(t)} \right\rangle,
\label{eq:Ed3}
\end{gather}
one can express the last term in \Eq{eq:Ed2} as
\begin{gather}
\favr{f^\alpha(t)\delta(w-u)}=-A^{\alpha\beta}\partial_{w^\gamma}
\left\langle \frac{\delta u^\gamma(t)}{\delta f^\beta(t)}\,\delta(w-u)\right\rangle.
\label{eq:Ed4}
\end{gather}
It follows from \Eq{eq:Ed0} that $\delta u^\alpha(t)/\delta f^\beta(t)=\delta^\alpha_\beta/2$ and therefore \Eq{eq:Ed2} for $F$ becomes an equation of the Fokker--Planck form,
\begin{gather}
\pd_t F + \pd_{w^\alpha}[g^\alpha(t,w)F]=\frac{1}{2}\,\pd_{w^\alpha}(A^{\alpha\beta}\partial_{w^\beta}F).
\label{eq:Ed5}
\end{gather}
Unlike \Eq{eq:Fcont}, this equation does not allow a simple Schr\"odinger representation. However, since \Eq{eq:Ed5} is linear, it can be solved using the general methods described in \Sec{sec:linearnonherm}. This can be used, for example, for studying homogeneous Navier--Stokes turbulence \cite{book:mccomb, ref:edwards64}.

%--------------------------------------
\subsection{Discussion}
\label{sec:subdiscussion}

To recap the above findings, the most interesting and relevant plasma problems are not immediately suited for the traditional QC architecture, which is a fit mainly for linear Schr\"odinger equations with Hermitian Hamiltonians. In order to map plasma problems to this architecture, it appears necessary to extend the configuration space~$\mc{C}$ (which is also how it is done in the recent \Refs{ref:joseph20, tex:engel20, tex:liu20, tex:lloyd20}.) Handling non-Hermiticity requires that the system size be only doubled (\Sec{sec:linearnonherm}), which is tolerable; however, nonlinearity presents a bigger challenge. 

Here, we have proposed a universal approach that allows for an arbitrary nonlinearity and dissipation. The idea is to encode the information about a dynamical system into a state vector that determines the probability of the system to be in a given part of $\mc{C}$ (\Sec{sec:general}). The dynamics of this vector is linear and unitary, so it can be naturally mapped to the QC architecture. Simulations for multiple initial conditions can be performed in parallel; this can be beneficial, for example, in optimization problems, where multiple initial guesses need to be processed for finding the global minimum. 

The required computational resources, or the number of qubits $N$, scale in our approach logarithmically with the required resolution and linearly with the number of degrees of freedom. That makes our approach particularly attractive for nonlinear-ODE solvers, where the number of degrees of freedom is not too large and the corresponding quantum Hamiltonians are sparse. Then, the corresponding run time scales linearly with~$N$. This scaling is fundamentally different from that in the commonly cited \Ref{tex:leyton08}, where the independent variable is encoded in the amplitude of the state function directly rather than through the probability amplitude. As a result, the run time and the required number of qubits in \Ref{tex:leyton08} scale logarithmically with the number of degrees of freedom but exponentially with the number of steps (\Sec{sec:prelim}). Also notably, an algorithm similar to that in \Ref{tex:leyton08} has been proposed recently for quantum optimization of polynomial functionals and exhibits similar scalings \cite{tex:rebentrost18}. The exponential scaling appears unavoidable for all algorithms of this type; hence, they are practically applicable only when the required number of steps is small. This, perhaps, rules them out as ODE solvers for plasma simulations. As a side note, though, such algorithms might be suitable for solving optimization problems in plasma physics. This is seen from the following example.

Let us consider the problem of magnetic-field optimization for the recently proposed permanent-magnet stellarator~\cite{ref:helander20}. The problem consists of finding the locations $\vec{x}_i$ and the dipole moments $\vec{m}_i$ of permanent magnets (subject to the engineering constraints) that produce a certain ``target'' field $\vec{B}_T$ within a prescribed volume. To uniquely specify such field, it is sufficient to specify the normal component of $\vec{B}_T$ on the volume boundary $\mc{S}$. Then, the problem can be reduced to minimizing the objective function $\mc{I}(\vec{y}) \doteq \int_{\mc{S}}[({\bf B}-{\bf B}_T) \cdot {\bf n}]^2 \dd \vec{s}$, where $\vec{B}$ is the actual field produced by the magnets, $\vec{n}$ is the unit vector field normal to $\mc{S}$, and $\vec{y} \doteq \{\vec{x}_i, \vec{m}_i\}$ is the array of all independent variables. The magnetic field can be approximated with a nonlinear polynomial function $\mc{I}(\vec{y})$. Then, the standard approach to optimizing~$\mc{I}$ is to reduce the set of free parameters~$\vec{y}$ to some smaller set~$\vec{z}$ that has the biggest impact on plasma performance; however, doing so limits the degree of optimization. A quantum algorithm potentially can do better, since it can handle much more degrees of freedom, perhaps, even the actual $\vec{y}$. The exponential scaling with the number of steps, which is the main bottleneck of the algorithm in \Ref{tex:rebentrost18}, is not a problem here, because the anticipated number of the iteration steps is not large (assuming a good initial guess is available). Therefore, by using the algorithm from \Ref{tex:rebentrost18}, one might be able to find a more optimal field configuration and thus improve plasma performance. 

%%%%%%%%%%%%%%%%%%%%%%%%%%%%%%%%%%%%%%%
\section{Eigenmodes and plasma stability}
\label{sec:eigen}

Another class of numerical plasma-physics problems for which QC can be useful is the problem of finding global linear eigenmodes and their frequencies $\omega$ \cite{ref:parker20}. Such problems emerge naturally, for example, in the context of MHD stability of fusion devices. As commonly known, eigenmodes of a static plasma governed by ideal MHD satisfy \cite{book:friedberg}  
\begin{gather}\label{eq:mhd}
-\omega^2\rho \, \xi^a = \oper{\mc{F}}^a{}_b\xi^b,
\end{gather}
where $\xi$ is a vector field that characterizes plasma displacement from a given equilibrium, $\rho$ is the equilibrium density, and $\oper{\mc{F}}$ is a linear operator that is Hermitian under the inner product $\braket{\xi|\eta} = \int \xi_a^*\eta^a\,\dd\vec{x}$; accordingly, all $\omega^2$ are real, while $\omega$ can be real or imaginary. Equation \eq{eq:mhd} can be rewritten as follows:
\begin{gather}\label{eq:mhd2}
\oper{H}\psi = \lambda \psi, 
\quad
\psi \doteq \rho^{1/2}\xi,
\quad
\oper{H} \doteq \rho^{-1/2} \oper{\mc{F}} \rho^{-1/2},
\end{gather}
and $\lambda \doteq -\omega^2$ are real. Since $\oper{\mc{F}}$ is Hermitian, so is $\oper{H}$. Then, \Eq{eq:mhd2} belongs to the class of problems that yield to known efficient quantum algorithms.\footnote{Notably, there also exist quantum algorithms for calculating (complex) eigenvalues of non-Hermitian operators \cite{ref:daskin13}. However, these algorithms are considerably less efficient.} One of them is the earliest quantum eigensolver \cite{ref:abrams99}, which is related to the HHL algorithm mentioned earlier. Another option is a hybrid quantum--classical method \cite{ref:peruzzo14, ref:mcclean16}, which can be efficient provided that: (i) $\oper{H}$ can be split into a polynomial sum of few-qubits operators, $\oper{H} = \sum_n \oper{H}_n$, and (ii) one can prepare ``ansatz'' quantum states on demand that cover the relevant part of Hilbert space with a given finite list of classical parameters. This hybrid method is briefly described as follows.

First, one calculates the ``ground state'', which corresponds to the smallest eigenvalue $\lambda$. To do that, one starts by preparing an ansatz state $\Psi$ with some trial parameters and calculates $H_n \doteq \braket{\Psi|\oper{H}_n|\Psi}$ on a quantum computer. Then, one feeds the results into a classical computer. The latter calculates $H \doteq \braket{\Psi|\oper{H}|\Psi}$ by summing up $H_n$ and then applies an iterative classical algorithm to adjust the parameters of the ansatz state such that $H$ be minimized. The resulting eigenstate is termed $\Psi_0$, and the corresponding eigenvalue is found as $\lambda_0 = \braket{\Psi_0|\oper{H}|\Psi_0}$. Next, one similarly minimizes $H$ in the subspace of vectors orthogonal to $\Psi_0$ and obtains the next eigenstate $\Psi_1$ and the corresponding eigenvalue $\lambda_1 = \braket{\Psi_1|\oper{H}|\Psi_1}$, and so on. Alternatively, the eigenvalues $\lambda$ of $\smash{\oper{H}}$ can be found as the local minima of the functional $\smash{\braket{\Psi| (\oper{H}-\lambda \unitoper)^2|\Psi}}$.

This algorithm allows one to find both real and imaginary eigenfrequencies $\omega = \sqrt{-\lambda}$ and thus explore plasma stability within ideal MHD. The quantum computer is used as a co-processor whose role is to efficiently calculate the matrix elements $H_n$, in which it can significantly outperform a classical computer \cite{ref:peruzzo14, ref:mcclean16}. Also note that the hybrid method imposes less strict requirements on the hardware. Each quantum calculation evaluates only a single matrix element, so the coherence time can be much smaller than that needed for solving the whole problem solely on a quantum computer.

%%%%%%%%%%%%%%%%%%%%%%%%%%%%%%%%%%%%%%%
\section{Variational approach to nonlinear simulations}
\label{sec:varia}

The hybrid quantum--classical variational approach can also be used for general simulations, including simulations of dissipative and nonlinear systems, as proposed in \Ref{ref:lubasch20}. Like in the previous case, one works with an ansatz quantum state $\phi(\theta)$ that is prepared on demand for a given finite list of classical parameters $\theta$. Suppose that at some time~$t_n$, one has $\phi(\theta_n) \approx  \psi(t_n)$, which is an approximation to a true solution $\psi(t_n)$ of the general type system $\partial_t \psi=\oper{O}(t)\psi$ at time $t_n$. At the next time step $t_{n+1} = t_n + \tau$, the true solution is given by
\begin{gather}
\psi(t_{n+1}) \approx 
\big[\,\unitoper + \tau \oper{O}(t_n)\big]\psi(t_n)
\approx  \big[\,\unitoper + \tau \oper{O}(t_n)\big]\phi(\theta_n).
\end{gather}
For linear systems, the operator $\oper{O}$ is prescribed and thus known at all times. For systems with polynomial nonlinearity, all nonlinear terms in $\oper{O}(t_n)\phi(t_n)$ can be evaluated using the projection method of \Ref{tex:leyton08} (see \Sec{sec:prelim}), since multiple copies of $\phi(\theta_n)$ can be constructed in parallel at any given time without restarting the simulation. Therefore, one can use a quantum computer to efficiently evaluate the ``cost function''
\begin{multline}
C(\theta_{n+1}) \doteq 
||\phi(\theta_{n+1})-\psi(t_{n+1})||^2
\\
\approx ||\phi(\theta_{n+1}) - [I +\tau \oper{O}(t_n)\phi(\theta_n)]\,||^2
\end{multline}
for any $\theta_{n+1}$. Then, one can efficiently find $\theta_{n+1}$ that minimizes the quantity $C(\theta_{n+1})$ using a classical computer. This amounts to finding $\phi(\theta_{n+1})$ that is maximally close to the true solution $\psi(t_{n+1})$; in other words, the system is integrated from $t_n$ to $t_{n+1}$. 

This process can be iterated from the initial moment of time, when $\psi$ is given, for any number of steps. Much like in \Sec{sec:eigen}, the role of the quantum computer here is limited to evaluating the cost function, while the optimization is done using a classical computer, which makes the scheme hybrid. The potential disadvantage of this method is that the simulation accuracy strongly depends on how closely the ansatz $\phi$ can approximate the true solution $\psi$. However, using an ansatz also has important advantages. Since each $\phi(\theta_n)$ is constructed independently for given $\theta_n$, such algorithm does not require exponentially many copies of $\phi$, unlike the method in \Ref{tex:leyton08}. Also, the ansatz-based method does not require extension of the configuration space assumed in \Sec{sec:nonlin}. This can be useful for solving nonlinear partial differential equations, whose configuration space on a grid is large. For example, \Ref{ref:lubasch20} describes application of the hybrid variational algorithm to solving a nonlinear Schr\"odinger equation, which is a common model in theory of nonlinear plasma waves.

%%%%%%%%%%%%%%%%%%%%%%%%%%%%%%%%%%%%%%%
\section{Conclusions}
\label{sec:conc}

Unlike quantum-mechanical systems that, in principle, can be mapped to the QC architecture more or less straightforwardly, modeling classical systems with quantum computers is challenging even at the conceptual level. Here, we report a preliminary exploration of the long-term opportunities and likely obstacles in this area. First, we show that many plasma-wave problems are naturally representable in a quantumlike form and thus are naturally fit for quantum computers. Second, we consider more general plasma problems that include non-Hermitian dynamics (instabilities,  irreversible dissipation) and nonlinearities. We show that by extending the configuration space, such systems can also be represented in a quantumlike form and thus can be simulated with quantum computers too, albeit that requires more computational resources compared to the first case. Third, we outline potential applications of hybrid quantum--classical computers, which include analysis of global eigenmodes and also an alternative approach to nonlinear simulations.

The work was supported by the U.S. DOE through Contract No.~DE-AC02-09CH11466. The authors also thank Stuart Hudson for valuable input.

%%%%%%%%%%%%%%%%%%%%%%%%%%%%%%%%%%%%%%%
\appendix

%%%%%%%%%%%%%%%%%%%%%%%%%%%%%%%%%%%%%%%
\section{Landau--Lifshitz--Gilbert equation}
\label{app:llg}

Consider a Schr\"odinger equation
\begin{gather}\label{eq:Psi}
\ii \dot{\Psi} = \oper{H} \Psi,
\end{gather}
where the Hamiltonian $\oper{H}$ is a non-Hermitian matrix. Let us introduce $a \doteq |\Psi|$ and $\psi \doteq \Psi/a$, so $|\psi| = 1$. Then,
\begin{gather}
\dot{a} = (\psi^{\dag}\oper{H}_A\psi)a,
\end{gather}
and one obtains the following \textit{nonlinear} equation for $\psi$:
\begin{gather}\label{eq:nlpsi}
\dot{\psi}=- \ii \oper{H}_H \psi + \oper{H}_A\psi - (\psi^{\dag} \oper{H}_A\psi)\psi.
\end{gather}
By reversing the argument, one can say that the nonlinear system \eq{eq:nlpsi} can be mapped to the linear system \eq{eq:Psi} by adding one extra dimension (the norm of $\psi$); then, it can be simulated as described in \Sec{sec:linearnonherm}.

The trace of $\oper{H}$ can be removed from the equation by a straightforward variable transformation, so we can assume that $\oper{H}$ is traceless without loss of generality. Then, $\oper{H}$ can be decomposed in the basis of the (Hermitian) generators $T_n$ of SU$(d)$, where $d \doteq \dim \psi$ \cite{foot:haber19}. For example, let us consider $d = 2$, which which case $T_n = \sigma_n/2$, where $\sigma_n$ are Pauli matrices. Let us assume the decomposition
\begin{gather}
\oper{H}_H = \frac{1}{2}\,\vec{B} \cdot \vec{\sigma},
\quad
\oper{H}_A=\frac{1}{2}\,\vec{R} \cdot \vec{\sigma},
\end{gather}
where $\vec{B}$ and $\vec{R}$ are real three-dimensional vectors and $\vec{\sigma} = \lbrace \sigma_1, \sigma_2, \sigma_3 \rbrace$ is a vector that has the Pauli matrices as its components. Then, \Eq{eq:nlpsi} becomes
\begin{gather}
\dot{\psi}
=- \frac{\ii}{2}(\vec{B} \cdot \vec{\sigma}) \psi 
+\frac{1}{2}\,(\vec{R} \cdot \vec{\sigma})\psi 
-\frac{1}{2}\,[\psi^{\dag}(\vec{R} \cdot \vec{\sigma})\psi]\psi.
\end{gather}
Let us also consider the effective ``spin'' vector
\begin{gather}
\vec{S} \doteq \psi^{\dag}\vec{\sigma} \psi.
\end{gather}
Then, a straightforward calculation shows that
\begin{gather}\label{eq:SLLG}
\dot{\vec{S}} = \vec{B} \times \vec{S} + \vec{R} - \vec{S} (\vec{R} \cdot \vec{S}).
\end{gather}
Note that just like \Eq{eq:nlpsi}, this nonlinear equation is in fact a representation of the linear system \eq{eq:Psi}.

Let us also consider the special case when $\vec{R} = \alpha \vec{B}$, where $\alpha$ is a scalar coefficient. Then, \Eq{eq:SLLG} is simply the Landau--Lifshitz--Gilbert equation, which is commonly used in theory of ferromagnetism \cite{ref:skrotskii84}. In this case, the evolution operator $\oper{G}$ of \Eq{eq:Psi}, defined via $\Psi(t) = \oper{G}\Psi(0)$, allows an explicit polar decomposition $\oper{G} = \oper{\mc{P}}\oper{\mc{U}}$, where
\begin{gather}
\oper{\mc{P}} 
= \ee^{\oper{H}_A t}
= \unitoper \cosh (\alpha \Omega t)+ (\vec{b} \cdot \vec{\sigma} ) \sinh (\alpha \Omega t),
\\
\oper{\mc{U}} 
= \ee^{-\ii \oper{H}_H t} 
= \unitoper \cos (\Omega t)-\ii (\vec{b} \cdot \vec{\sigma}) \sin(\Omega t),
\end{gather}
with $\Omega \doteq |\vec{B}|/2$ and $\vec{b} \doteq \vec{B}/|\vec{B}|$. Interestingly, the resulting dissipative (or nonlinear) model is solvable via QHS. Specifically, the spin vector can be expressed as follows:
\begin{gather}
\vec{S} 
=\frac{\Psi^{\dag} \vec{\sigma} \Psi}{\Psi^{\dag}\Psi}
=\frac{\xi^{\dag} \oper{\mc{P}} \vec{\sigma}\oper{\mc{P}} \xi}{\xi^{\dag}\widehat{\mc{P}}^2 \xi},
\end{gather}
where $\xi \doteq \oper{\mc{U}} \Psi(0)$. This can also be simplified as
\begin{gather}
\vec{S} = (\xi^{\dag} V \vec{\sigma} V \xi)/(\xi^{\dag} V^2 \xi),
\\
V = \unitoper + (\vec{b}\cdot \vec{\sigma}) \tanh (\alpha \Omega t).
\end{gather}
The numerator and the denominator can be calculated separately using QHS, and then one can divide one over another using a classical computer.\\

%%%%%%%%%%%%%%%%%%%%%%%%%%%%%%%%%%%%%%%
\section{Generalized functions $\boldsymbol{\delta}$ and $\boldsymbol{\Delta}$}
\label{app:delta}

Here, we present a more rigorous definition of the ``function'' $\Delta$ that we have introduced in the main text symbolically as the square root of the Dirac delta function $\delta$. We start by revisiting the known definition of $\delta$ and then define $\Delta$ by analogy. 

In what follows, we limit our considerations to functions of a one-dimensional coordinate $x$ (the generalization to multiple dimensions is straightforward) and assume the standard definition of the inner product of two given functions $f$ and $g$,
\begin{gather}\label{eq:inner}
\braket{f|g} = \int f^*(x)g(x)\,\dd x.
\end{gather}
This defines a Hilbert space $\msf{H}$ where functions are vectors, or kets, $\ket{g}$. Covectors, or bras, are linear mappings on kets, $\bra{f} \doteq \int f^*(x)(\ldots)\,\dd x$, so \Eq{eq:inner} describes the application of $\bra{f}$ to~$\ket{g}$. Kets and bras are connected by bijection $\ket{f} \leftrightarrow f \leftrightarrow f^* \leftrightarrow \bra{f}$.

%--------------------------------------
\subsection{Generalized functions in the continuous space}

The delta ``function'' $\delta$ is a distribution \cite{tex:schwartz63} that implements the linear mapping $a \mapsto a(0)$. If $a$ is viewed as a ket $\ket{a}$, such mapping is by definition a bra $\bra{\delta}$ that~satisfies
\begin{gather}\label{eq:dp}
\delta: \kern 5pt \ket{a} \mapsto \braket{\delta|a} = a(0).
\end{gather}
If one formally treats $\delta$ as a (real) function, then \Eq{eq:dp} can be written as
\begin{gather}\label{eq:dint}
\int \delta(x) a(x)\,\dd x = a(0),
\end{gather}
from where it is seen that the delta function is even, $\delta(x) = \delta(-x)$. Such function can be modeled, for example, as a sufficiently narrow Gaussian with $\int \delta(x)\,\dd x = 1$.

Let us consider functions $\delta(x-\lambda)$ with different real $\lambda$ and denote the corresponding kets as $\ket{\delta_\lambda}$. Each such ket can is an eigenvector of the coordinate operator $\oper{x}$ corresponding to the eigenvalue $\lambda$; \ie $\oper{x}\ket{\delta_\lambda} = \lambda\ket{\delta_\lambda}$. Also, from \Eq{eq:dint}, one has
\begin{gather}\label{eq:qnorm}
\braket{\delta_{\lambda_1}|\delta_{\lambda_2}} 
= \int \delta(x-\lambda_1)\delta(x-\lambda_2)\,\dd x
= \delta(\lambda_1 - \lambda_2).
\end{gather}
This makes $\{\ket{\delta_\lambda}\}$ convenient as a basis for vectors in $\msf{H}$, because the corresponding coordinates $\braket{\delta_{\lambda}|a}$ of any finite vector $a$ are finite. However, $\ket{\delta_\lambda}$ have infinite norm and thus, strictly speaking, do not even belong to $\msf{H}$. Moreover, elements of a finite-norm $(n>1)$-rank tensor in the corresponding basis (made of tensor products of $\ket{\delta_\lambda}$ and $\bra{\delta_\lambda}$) are typically singular. This  motivates construction of ``generalized distributions'' that induce more suitable bases for such tensors and thus allow meaningful generalizations of \Eq{eq:dp} from the vector mapping to tensor mappings.

In particular, let us consider a linear mapping of some rank-2 tensor, specifically, some operator $\oper{A}$, to a (generally complex) number. Suppose this mapping is determined by some $\Delta$, which induces a dyadic $\ket{\Delta}\bra{\Delta}$:
\begin{gather}\label{eq:A}
\Delta: \kern 5pt \oper{A} \mapsto \braket{\Delta|\oper{A}|\Delta}.
\end{gather}
By analogy with $\ket{\delta}$, we require that $\ket{\Delta}$ be an eigenvector of $\oper{x}$ corresponding to the zero eigenvalue. By analogy with \Eq{eq:dint}, we also require that $\braket{\Delta|\oper{A}|\Delta}$ be finite if $\oper{A}$ has a finite norm; then, $\braket{\Delta|\Delta}$ must be finite too, and we choose it to be unity. This completely specifies $\ket{\Delta}$. Now suppose that $\oper{A}=A(\oper{x})$, where $A$ is a finite function.~Then,
\begin{gather}\label{eq:Dp}
\braket{\Delta|\oper{A}|\Delta} = A(0).
\end{gather}
If one formally treats $\Delta$ as a (real) function, then \Eq{eq:Dp} can be written as
\begin{gather}\label{eq:Dint}
\int \Delta^2(x) A(x)\,\dd x = A(0).
\end{gather}
Then, by comparing \Eq{eq:Dint} with \Eq{eq:dint}, one finds that $\Delta(x) = \sqrt{\delta(x)}$. If $\delta$ is modeled, say, by a narrow Gaussian, this defines $\Delta$ as another narrow Gaussian \cite{ref:craven85}.

Let us consider functions $\Delta(x-\lambda)$ with different real $\lambda$ and denote the corresponding kets as $\ket{\Delta_\lambda}$. Like $\ket{\delta_\lambda}$, they are mutually orthogonal eigenvectors of~$\oper{x}$,
\begin{gather}
\oper{x}\ket{\Delta_\lambda} = \lambda\ket{\Delta_\lambda}, 
\quad 
\braket{\Delta_\lambda|\Delta_\mu} = 0 
\kern 8pt {\rm for} \kern 8pt \lambda \ne \mu.
\end{gather}
However, unlike $\ket{\delta_\lambda}$, these vectors have unit norm, $\braket{\Delta_\lambda|\Delta_\lambda} = 1$.

%--------------------------------------
\subsection{Generalized functions on a grid}
\label{app:delta3}

On a grid with cell coordinates $\{x_n\}$ and cell size $q$, the functions $\delta$ and $\Delta$ can be represented as
\begin{gather}
\delta(x_n) = \left\lbrace
\begin{array}{cc}
0, & n \ne 0\\
q^{-1}, & n = 0
\end{array}
\right.
,
\\
\Delta(x_n) = \left\lbrace
\begin{array}{cc}
0, & n \ne 0\\
q^{-1/2}, & n = 0
\end{array}
\right.
,
\end{gather}
where the cell index $n = 0$ corresponds to $x_n=0$. In this case, both functions are finite and have finite norms: 
\begin{gather}
\braket{\delta_m|\delta_n} = q^{-1}\delta_{m,n}, 
\quad 
\braket{\Delta_m|\Delta_n} = \delta_{m,n},
\end{gather}
where $\delta_{m,n}$ is the Kronecker symbol. Hence, one can work with them like with any other finite-dimensional vectors.

%\gap\\
%\bibliography{main,my,qc,turbulence,foot} 

\end{document}